\documentclass[aps,prb,twocolumn,showpacs,showkeys,groupedaddress,floatfix,amsmath,amssymb]{revtex4}

\usepackage{graphicx}
\usepackage{dcolumn}
\usepackage{bm}

\begin{document}

\title{Stationary phase slip state in quasi-one-dimensional rings.}

\author{D. Y. Vodolazov}
\author{B. J. Baelus}
\author{F. M. Peeters}
\email{peeters@uia.ua.ac.be} \affiliation{Departement Natuurkunde,
Universiteit Antwerpen (UIA), Universiteitsplein 1,  B-2610
Antwerpen, Belgium}

%\date{\today}

\begin{abstract}
The nonuniform superconducting state in a ring in which the order
parameter vanishing at one point is studied. This state is
characterized by a jump of the phase by $\pi$ at the point where
the order parameter becomes zero. In uniform rings such a state is
a saddle-point state and consequently unstable. However, for
non-uniform rings with e.g. variations of geometrical or physical
parameters or with attached wires this state can be stabilized and
may be realized experimentally.
\end{abstract}

\pacs{74.60Ec, 74.20.De, 73.23.-b} \keywords{phase slip,
mesoscopic superconductivity}

\maketitle

In recent years \cite{Berger,Horane,Berger1} the existence of a
single-connected state in a ring was proposed theoretically. Such
a state implies that the relation between the phase $\phi$ of the
order parameter $\psi=fe^{{\rm i}\phi}$, the current density $j$
and the magnetic flux $\Phi$ through the ring (which follows from
the single valuedness of $\psi$ - see for example Ref.
\cite{Tinkham})
\begin{equation}
\oint \frac{j}{f^2}ds=2\pi n-\Phi,
\end{equation}
is no longer valid (the flux is expressed in $\Phi_0/2\pi$, the
current density in $j_0=c\Phi_0/4\pi^2\lambda^2\xi$, $\lambda$ is
the London penetration length, $\xi$ the coherence length and
$\Phi_0$ is the quantum of magnetic flux). The reason is that the
order parameter in such a single-connected state is zero at one
point. Moreover, it was claimed that under certain conditions,
i.e. radius of the ring less than $\xi$ and the flux through the
ring is about $(n+1/2)\Phi_0$ this state may become metastable in
some range of magnetic fields \cite{Horane}.

In this paper we revisited this problem and we will show that a
state where the order parameter vanishes in one point is still
double-connected and that Eq. (1) is valid for such a state. We
propose to call such a state a one-dimensional vortex state (ODV
state), because like for an ordinary two-dimensional Abrikosov
vortex, there is a jump in the phase of the order parameter of
$\pi$ at the point where $\psi=0$. In contrast to a
two-dimensional Abrikosov vortex the 1D vortex is an unstable
structure in an uniform ring. However, if some inhomogeneities are
present in the ring (defects, nonuniform thickness or nonuniform
width of the ring, attached superconducting wires) the ODV
structure can be stabilized and may consequently be realized
experimentally.

Consider a uniform ring with thickness $d \lesssim \lambda$ and
width $w \lesssim \xi$. In addition, let the radius of the ring
$R$ be much larger than $w$. Under these conditions we can neglect
the screening effects and the problem is reduced to a
one-dimensional one. The distribution of the current density $j$
and the order parameter $\psi$ of the system at a temperature not
far from the critical temperature $T_c$ is described by the 1D
Ginzburg-Landau equation (plus the condition ${\rm div} {\bf
j}=0$)
\begin{subequations}
\begin{equation}
 \frac{d^2f}{ds^2}+f(1-f^2-p^2)=0, %2a
\end{equation}
\begin{equation}
 \frac{dj}{ds}=\frac{d}{ds}f^2p=0, %2b
\end{equation}
\end{subequations}
where the gauge-invariant momentum $p=\nabla \phi -A$ is scaled in
units of $\Phi_0/(2\pi\xi)$, the length of the ring is $L=2\pi R$
and the circular coordinate $s$ is in units of the coherence
length $\xi$. In these units, the magnetic field is scaled in
units of the second critical field, $H_{c2}$, and the magnetic
flux in $\Phi_0/2\pi$.

The coupled Eqs. (2a) and (2b) have to be solved with the boundary
condition $\psi(-L/2)=\psi(L/2)$. We  use the method proposed in
\cite{Langer} (see also \cite{Zhang}). These equations have the
first integral
\begin{equation}
\frac{1}{2}\left(\frac{df}{ds}\right)^2+\frac{f^2}{2}-\frac{f^4}{4} %3
+\frac{j^2}{2f^2}=E.
\end{equation}
From a formal point of view, Eq. (3) is nothing else then the
condition of "energy", $(E)$, conservation for some "particle"
with "coordinate" $f$ and "momentum" $j$. The role of "time" is
played \cite{Langer} by the circular coordinate $s$ . In Fig. 1 we
show the dependence of the "potential energy" of this system
\begin{equation}
U(f)=\frac{f^2}{2}-\frac{f^4}{4}+\frac{j^2}{2f^2}, %4
\end{equation}
for different values of $j$. Possible solutions of Eqs. (2a,b) are
in the region where confined "trajectories" of our virtual
particle exist. This is possible for currents $0<j\leq j_c$
($j_c=\sqrt{4/27}j_0$ is the depairing current density ).
\begin{figure}[ht]
\includegraphics[width=0.48\textwidth]{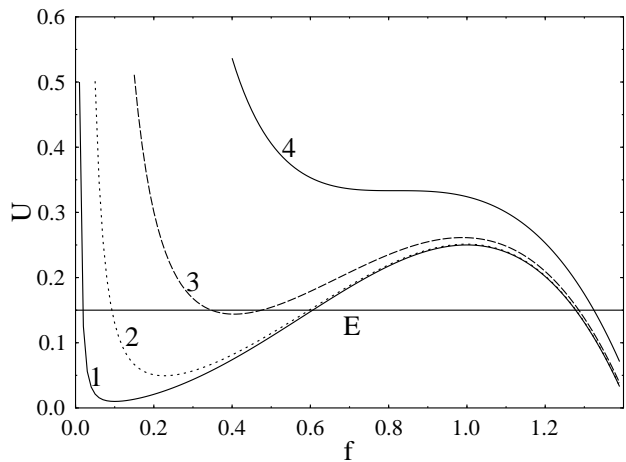}
\caption{Dependence of the "potential energy" $U(f)$ for different
values of the current $j$: 1- $j$=$0.01$, 2- $j$=$0.05$, 3-
$j$=$0.15$, 4- $j$=$j_c$=$\sqrt{4/27}$.}
\end{figure}
Using Eq. (3) we immediately can write the solution of Eqs. (2a,b)
for the ring
\begin{eqnarray}
\sqrt{2}s=\int_{t_0}^{t}\frac{dt}{\sqrt{(t-t_0)(t_1-t_0)(t_2-t_0)}}= \nonumber\\
\frac{2}{\sqrt{t_2-t_0}}F\left({\rm sin^{-1}}\sqrt{\frac{t-t_0} %5
{t_1-t_0}},\sqrt{\frac{t_1-t_0}{t_2-t_0}}\right),
\end{eqnarray}
where $t(s)=f^2(s)$, $F(\theta,m)$ is the elliptic integral of the
first kind and $t_0\leq t_1\leq t_2$ are the solutions of the
cubic equation
\begin{equation}
t^3-2t^2+4Et-2j^2=0. %6
\end{equation}
Using the boundary condition for $f$ we may conclude, that for a
given current there exist a maximum of three solutions. There are
two uniform solutions (in the points of the minimum $E=E_{min}$
and the maximum $E=E_{max}$ of the "potential energy"(4)) and one
nonuniform solution with energy $E_{min}<E<E_{max}$ which is
defined by the equation
\begin{equation}
L=\sqrt{\frac{8}{t_2-t_0}}K\left(\sqrt{\frac{t_1-t_0}{t_2-t_0}}\right), %7
\end{equation}
where $K(m)$ is the complete elliptic integral of the first kind.
Numerical analysis of Eq. (7) for $0<j<j_c$ shows that there is a
minimal value for the ring length $L_{min}$ for which there exists
a solution for this equation. When $j\to j_c$, $L_{min} \to
\infty$ and in the opposite limit $j \to 0$ one can show that
$L_{min} \to \pi$. The latter corresponds to a ring radius $R=1/2$
(or $\xi/2$ in dimensional units). For a radius $R>1/2$ metastable
states may exist and superconductivity is present for any value of
the magnetic flux (at least in our one-dimensional model)
\cite{Horane,Buzdin}.

In principle, Eqs. (5-7) define the nonuniform solution of Eqs.
(2a,b) for a ring. Unfortunately, even using the explicit Kordano
expressions for the roots of Eq. (6) the results are rather
complicated for arbitrary values of the current $j$. However a
tractable analytical solution is possible if we consider the case
of low currents $j \ll 1$ for which the roots of Eq. (6) are
simplified to
\begin{subequations}
\begin{equation}
 t_0 \simeq j^2/2E ,  %8a
\end{equation}
\begin{equation}
 t_1\simeq 1-\sqrt{1-4E} , %8b
\end{equation}
\begin{equation}
t_2\simeq 1+\sqrt{1-4E} . %8c
\end{equation}
\end{subequations}
After inserting these results into Eq. (7) we obtain the "energy".
For $L \gg \pi$ the "energy" $E\simeq 1/4$ is practically
independent of $L$ (e.g. for $L=8$ the difference $1/4-E(L=8)$ is
less than 0.002). Also in the limit $L-\pi \ll 1$ we found that
the "energy" $E$ is independent of $j$
\begin{equation}
E(L) \simeq \frac{1}{2}\left(\frac{L-\pi}{\pi/8}\right)^2. %9
\end{equation}
In the limit $j \ll 1$ we can also find \cite{comm1} the
dependence of $t(s)$ near the minimum point of $t(s)$
\begin{subequations}
\begin{eqnarray}
t(s)=2j^2+s^2/2, & L\gg \pi, s\ll 1 \label{a}%10a
\\
t(s)=j^2/2E+2Es^2, &  L-\pi \ll 1,  s\ll 1 \label{b} %10b
\end{eqnarray}
\end{subequations}
Using Eqs. (10a,b) it is easy to show, that in the limit $j \to 0$
the gauge-invariant momentum is given by
\begin{equation}
p(s)=j/t(s)=\frac{j}{|j|}(\pi+2\pi m)\delta(s), %11
\end{equation}
where $\delta(s)$ is the Dirac function and $m$ is a integer. As a
result the integral
\begin{eqnarray}
\lim\limits_{j\to 0}\oint \frac{j}{f^2}ds=\lim\limits_{j \to
0}\oint p(s)ds= \nonumber
\\
=\phi(+\epsilon)-\phi(-\epsilon)=\Delta \phi=\pm %12
(\pi+2\pi m),
\end{eqnarray}
for arbitrary ring size \cite{comm2}. Combining Eq. (12) with Eq.
(1) it is easy to show that if $\Phi=(k+1/2)\Phi_0$ ($k$ is an
integer) a solution of Eq. (2a) can be found which vanishes at one
point.

In previous \cite{Berger,Horane,Berger1} studies of this state Eq.
(11) was not taken into account (only the absolute value of the
order parameter was found). Furthermore, in Ref. \cite{Horane} it
was claimed that for small rings there is a magnetic field region
in which such a state exist and is metastable. But we find that in
this region this state even does not exist (except in one point)
because Eq. (1) is not fulfilled!

The distribution of the order parameter as obtained in the present
paper may also be applied to a ring in which some part of the ring
consists of a normal metal. Then at the boundary between the
normal metal and the superconductor the condition $\psi \simeq 0$
is fulfilled (if the normal part is longer than $\xi$) and the
distribution of the density $|\psi(s)|^2$ coincides with the one
obtained in Ref. \cite{Horane} (but the phase $\phi(s)$ will be
different for our previous ring geometry). Besides we cannot call
such a state single-connected as in Ref.
\cite{Berger,Horane,Berger1} because the relation (1) is valid
even for this case. Therefore, it is better to call this state a
one-dimensional vortex state because like the two-dimensional
Abrikosov vortex there is a point where $|\psi|=0$ and the phase
of the order parameter exhibits a jump equal to $k\pi$ (as the
Abrikosov vortex with orbital momentum $2\pi k$). A numerical
calculation shows that, as in the case for an ordinary Abrikosov
vortex in superconductors where such a state is stable, the phase
jump is always equal to $\pi$, i.e. $k=1$.
\begin{figure}[h]
\includegraphics[width=0.5\textwidth]{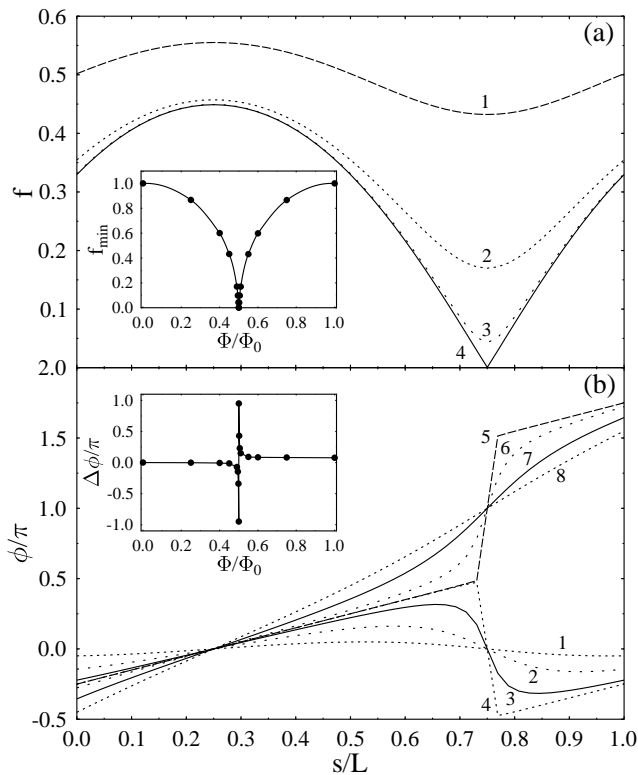}
\caption{Distribution of the absolute value (a) and phase (b) of
the order parameter at different values of the magnetic flux: 1-
$\Phi$=$0.45\Phi_0$, 2- $\Phi$=$0.49\Phi_0$, 3-
$\Phi$=$0.498\Phi_0$, 4- $\Phi$=$(0.5-0)\Phi_0$, 5-
$\Phi$=$(0.5+0)\Phi_0$, 6- $\Phi$=$0.502\Phi_0$, 7-
$\Phi$=$0.51\Phi_0$, 8- $\Phi$=$0.55\Phi_0$. Note that the order
parameter for $\Phi=\Phi_0/2+\alpha$ is the same as for
$\Phi=\Phi_0/2-\alpha$ when $\alpha<\Phi_0/2$. In the insets (a)
$f_{min}$ and (b) the phase difference $\Delta \phi$ near the
point $s/L=3/4$ are shown as function of $\Phi$.}
\end{figure}

Our numerical analysis of the time-dependent Ginzburg-Landau
equations showed that this state is completely unstable for a
uniform ring. However for rings with nonuniform width(thickness)
or attached wires a nonuniform distribution of the order parameter
becomes possible \cite{Berger,Berger1,Fink1,Fink2} and for small
rings with $L \sim \pi$ this state is realized in practice
(besides for the case of a ring, a ODV state can also exist in the
Wheatstone bridge - see Refs. \cite{Fink3,Ammann}). In Fig. 2 the
distribution of the absolute value and the phase of the order
parameter is shown for different values of the magnetic field for
a nonuniform ring \cite{comm3}. We used the following model where
the width of the ring was varied as
\begin{equation}
w(s)=1+w_0{\rm sin}\left(\frac{2\pi s}{L}\right), %13
\end{equation}
The parameters used are: $w_0=0.1$ and $L=3.25$. It is seen that
with increasing magnetic flux the magnitude of the order parameter
$f$ decreases in the thinner part of the ring (at $s/L=3/4$) and
becomes zero when $\Phi/\Phi_0=1/2$. In Ref. \cite{Berger1} it was
found that the ODV state is only possible at $\Phi=(n+1/2)\Phi_0$.
The reason is now clear - only at this values of the magnetic flux
the phase difference near the point where the order parameter is
zero will be compensated by the term $2\pi n-\Phi$ and the current
density will be equal to zero. It is interesting to note that the
current density is also zero for $\Phi=n\Phi_0$ even in the case
of a nonuniform ring, but for values of the flux where the ODV
state does not exist. At these values of the magnetic flux the
order parameter is uniform along the ring and the term $2\pi n$ is
completely compensated by the term $\Phi$ in Eq. (1).

\begin{figure}[hbtp]
\includegraphics[width=0.48\textwidth]{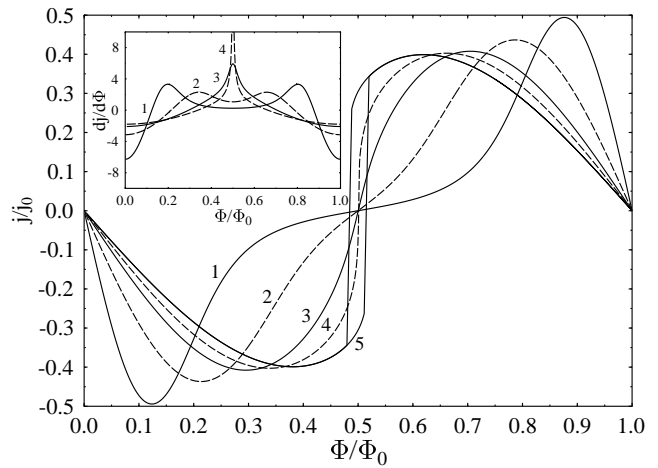}
\caption{The current in the ring (with attached wire of length
$5\xi$) as function of the flux. The different curves are for
different circumference of the ring: 1- $L$=1, 2- $L$=2, 3- $L$=3,
4- $L$=3.5, 5- $L$=4. For $L$=4 the ODV state does not exist in
the ring and hysteresis appears. In the inset the dependence of
$dj/d\Phi$ on the flux $\Phi$ is shown for rings with lengths
$L$=1, 2, 3, 3.5 (curves 1,2,3,4 respectively). Note that there
are two maxima in the range $(0,1)$. With increasing length the
two maxima merge into one and when one approaches the critical
length $dj/d\Phi$ diverges at $\Phi=\Phi_0/2$.}
\end{figure}

Above we showed that the ODV state can be realized by varying the
geometrical parameters of the ring. But there are two alternative
approaches to realize the ODV state in the ring experimentally.
First, it is possible to include an other phase in the ring. This
leads to the appearance of a weak link in the sample and if the
radius of the ring is less than some critical value (about $\xi$)
it also leads to the existence of the ODV state in the ring at
$\Phi=(n+1/2)\Phi_0$. We modelled this situation by introducing an
additional term $\rho(s)f$ in Eq. (2a), and, as an example, we
took $\rho(s)=-\alpha \delta(s)$ and $\alpha=1$ (it should be
noted that qualitatively the results do not depend on the specific
value of $\alpha$). We will not present the numerical results of
the modified Ginzburg-Landau equations but we found that they are
qualitatively similar to the behavior of $f$ and $\phi$ shown in
Fig. 2. When approaching the flux $\Phi=(n+1/2)\Phi_0$ the order
parameter reaches zero in the defect point and a jump in the phase
equal to $\pi$ occurs. Secondly, such a ODV state should also
appear in small rings with attached wire(s). As was shown in Refs.
\cite{Fink1,Fink2} an attached wire leads to a nonuniform
distribution of the order parameter in the ring. In Fig. 3 the
dependence of the current $j$ in such a ring on the flux through
the ring is shown (see also Ref. \cite{Fink2}). As for nonuniform
rings, where the parameters are chosen such that the ODV state may
exist, there is no hysteresis in such a system \cite{Berger1}.
\begin{figure}[t]
\includegraphics[width=0.48\textwidth]{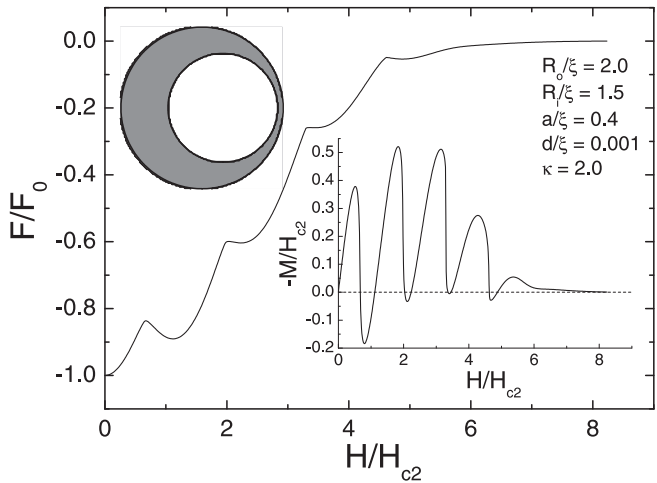}
\caption{The Gibbs free energy and the magnetization (right lower
inset) of the asymmetric ring (upper inset) as function of the
applied magnetic field.}
\end{figure}
Small inhomogeneities in rings gives us the unique possibility to
study states close to the ODV state because there is no sharp
transition from the state with finite order parameter to a state
with vanishing order parameter at one point (see Fig. 2). The
latter is, in some sense, a 'frozen' slip phase state. A small
variation of the flux through the ring near $\Phi=\Phi_0/2$ leads
to a change of $\Delta \phi$ by $2\pi$ (from $-\pi$ to $+\pi$ with
a concomitant change of the current from $-0$ to $+0$ - see Fig.
2(b)). As a result an additional phase circulation $\oint \nabla
\phi ds$=$2\pi$ appears in the system, not because the magnetic
flux $\Phi$ is changed, but because the term $\oint j/f^2 ds$
changes from $-\pi$ to $+\pi$ (see Eq. (1)). When such a jump
occurs $\nabla \phi$ does not change in the ring, except near the
point where $f=0$, and as a result the current density in the
system changes continuously. In the usual case transitions of the
vorticity from a state with phase circulation $2\pi n$ to a state
with phase circulation $2\pi (n+1)$ leads to a jump in the current
{\it everywhere} in the ring because $\oint \nabla \phi ds$
changes by $2\pi$ and the order parameter and the current density
are finite {\it everywhere} in the system.

To supplement the above study we made also a numerical study of a
nonuniform ring of finite width and thickness by implementing our
previous finite difference solution of the coupled nonlinear
Ginzburg-Landau equations \cite{Baelus1}. As an example we took
the following parameters: outer radius of the ring $R_o=2\xi$,
radius of the hole $R_i=1.5\xi$, displacement of the hole from the
center $a=0.4\xi$, ring thickness $d=0.001\xi$ and Ginzburg-Landau
parameter $\kappa=2$.

In Fig. 4 the dependence of the Gibbs free energy $F$ and the
magnetization $M=-\partial F/\partial H$ of this system on the
magnetic field are shown. As in the case of our one-dimensional
ring these dependencies are reversible and there are magnetic
fields at which the magnetization is equal to zero (in the points
of local maximum and minimum of the free energy). We found that
the distribution of the order parameter and the phase in the ring
(see inset of Fig. 5) is similar to analogical distributions for
the above one-dimensional ring (Fig. 2) at low magnetic fields.
But different with the one-dimensional system the phase
circulation increase of $2\pi$ (between the points (2) and (3) in
Fig. 5) does not occur at the magnetic field value where the free
energy has a local maximum (and zero magnetization). With
decreasing width of the ring this point shifts towards the local
maximum in the free energy.
\begin{figure}[t]
\includegraphics[width=0.35\textwidth,angle=-90]{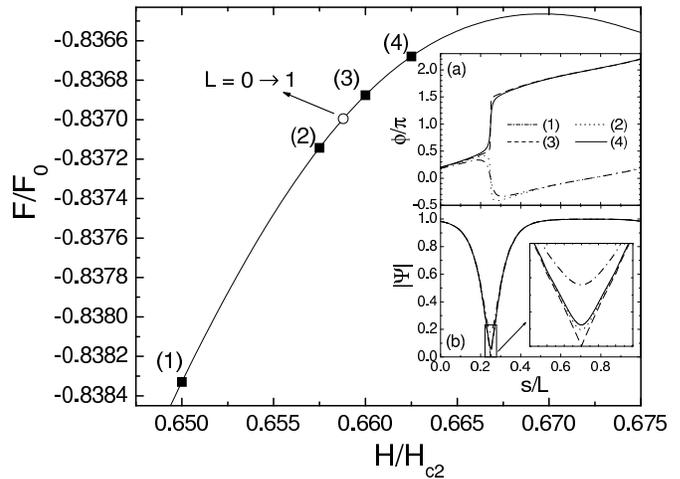}
\caption{The Gibbs free energy $F(H)$ near the first maximum. In
the inset the phase and the order parameter distribution at
different values of the applied magnetic field (indicated by the
squares on the $F(H)$ curve) are shown. The phase was calculated
along the outer perimeter of the ring. The open circle on $F(H)$
indicates the position at which the vorticity increases from 0 to
1.}
\end{figure}
It is interesting to note that for the system corresponding to
Fig. 4 the vortex enters through the thinnest part of the ring in
the case of the first three maxima in the free energy while for
the highest magnetic field maxima (i.e. $H/H_{c2}\simeq 4.6$) the
vortex enters through the thickest part of the ring (see Fig. 6).
Because the width of thicker part of the ring ($=0.9\xi$) is
considerable larger than the thinner part ($=0.1\xi$) and is of
order $\xi$ the ODV for $H/H_{c2}\simeq 4.6$ becomes the usual
two-dimensional vortex. As a result the order parameter is equal
to zero only in one point along the radial coordinate and the
circulation of the phase of the order parameter now is also a
function of this coordinate (see Fig. 6). When the vortex
enters/exits the ring at low magnetic fields there is also a slow
dependence of the order parameter on the radial coordinate along
the thinnest part of the ring. Therefore, we can conclude that in
a ring with finite width the one-dimensional vortex is transformed
into the usual two-dimensional vortex.

\begin{figure}[h]
\includegraphics[width=0.36\textwidth,angle=-90]{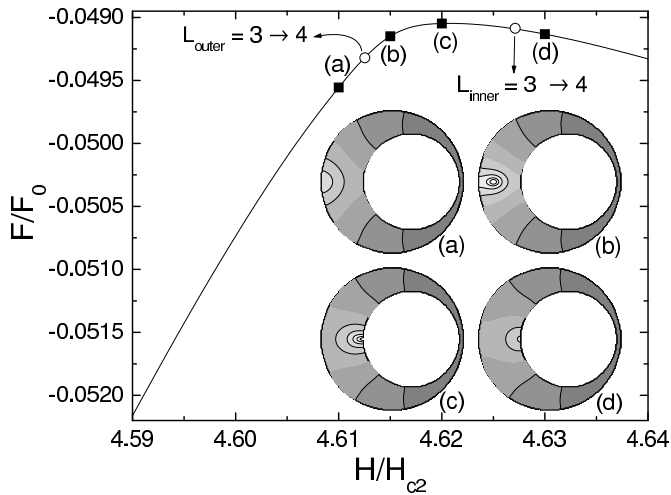}
\caption{The same as Fig. 5 but now for the last maximum in Fig.
4. In the inset a contourplot of the order parameter is shown for
different values of the magnetic field which are given by the
squares in the main figure. $L_{outer}$ is the vorticity as
calculated along the outer perimeter of the ring and $L_{inner}$
is the vorticity calculated along the perimeter of the hole.}
\end{figure}

The process of the appearance of a stable vortex in a ring with
finite width is similar to the creation of a phase slip in wires
of finite width in the presence of a transport current. In the
latter case the distribution of the gauge-invariant momentum is
not exactly uniform over the wire width. The maximum value of $p$
is obtained at the edges and as a result the order parameter
vanishes first in these points. When a phase slip is created the
distribution of the order parameter is not uniform over the width.
With decreasing wire width this nonuniformity decreases but it
will be uniform, strictly speaking, only in the limit $w \to 0$.
With increasing wire width the phase slip transforms to the
ordinary process of vortex/antivortex pair nucleating at the
edges, penetrating deep into the superconductor and annihilating.
In nonuniform rings the distribution of $p$ over the width is
nonuniform and in contrast to wires with transport current it is
nonsymmetric with respect to the ring width. Suppression of the
order parameter first occurs only on the external side (when
increasing the magnetic field) or on the internal side (when
decreasing the magnetic field). As in the case of an
one-dimensional ring we may call this state a stable ("frozen")
phase slip state if the width of the ring, where the vortex
penetrates, is less than $\xi$.

The ODV state of a nonuniform ring is very similar to the
saddle-point state as found in Ref. \cite{Baelus1} in case of an
uniform ring of finite width and in Ref. \cite{Schweigert} for the
case of a disk. Note that for an uniform ring with zero width the
nonuniform solution Eq. (6) corresponds to a saddle-point point of
$F(\psi)$. In Fig. 7 the gradual decrease of the hysteresis with
increasing displacement $a$ of the hole from the center of the
ring is shown. It is seen that with increasing $a$ the region
where metastable states exist decreases and ultimately vanishes
for some critical displacement $a_c$. The saddle point states
(only shown for $a=0$, by the thin full curve) approaches the
stable phase slip state when $a=a_c$.
\begin{figure}[t]
\includegraphics[width=0.36\textwidth,angle=-90]{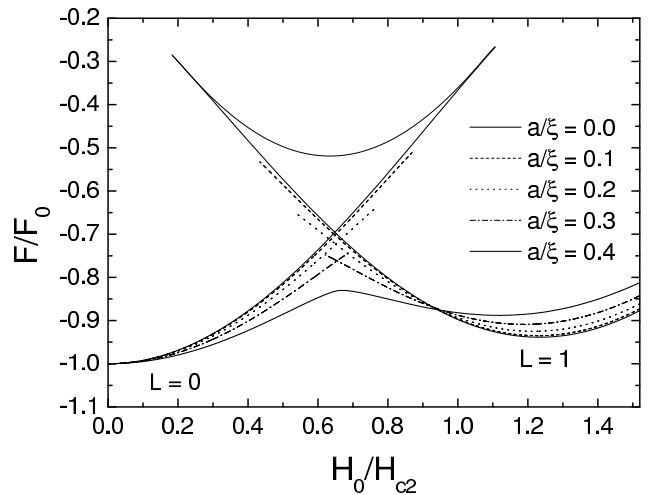}
\caption{The Gibbs free energy $F(H)$ at low magnetic fields for
different displacements $a$ of the hole from the center of the
ring. The thin solid curve (for $a=0$) corresponds to the saddle
point state.}
\end{figure}
In Ref. \cite{Baelus2} also nonuniform superconducting rings of
finite width were investigated (but with larger ring radius). It
was found that for low magnetic fields $F(H)$ is irreversible but
for high magnetic fields the dependence $F(H)$ was reversible. In
high fields the transition from a state with vorticity one to
another state occured through the same scenario as discussed
above. A similar state was also considered in Ref. \cite{Berger3}
in the framework of the linearized Ginzburg-Landau equations for a
nonuniform ring of finite width. They found that in such a system
the vortex may be stable in some magnetic field region. In low
magnetic field this vortex enter through the thinner part of the
ring and at high magnetic fields through the thicker part. Our
calculations generalizes this result to lower temperatures (i.e.
where the nonlinear Ginzburg-Landau equations have to be used) and
with inclusion of the non-zero demagnetization factor of the ring.
We found the dependence of the Gibbs free energy (and
magnetization) of this sample on the applied magnetic field. Our
results also allowed to compare the results of the one-dimensional
model with the full two-dimensional one.

The ODV (or stable phase slip) state may be observed by magnetic
experiments. Magnetic susceptibility is proportional \cite{Zhang}
to $dj/d\Phi$ and the ODV state exhibits some characteristic
peculiarities as was shown in the inset of Fig. 3. Magnetization
$M$ is proportional to the current $j$ and hence $M(H)$ is
reversible, for samples where the ODV state exists (see Fig. 3 and
inset of Fig. 5). Furthermore $M=0$ at $\Phi\simeq (n+1/2)\Phi_0$.
Alternatively, because at $\Phi=(n+1/2)\Phi_0$ there is a point in
the ring where the order parameter is equal to zero, this state
may be found by transport measurements. For example in Ref.
\cite{Liu} the dependence of the resistance of a hollow cylinder
with radius of order $\xi$ was studied at temperature $T<T_c$ far
from $T_c$. At $\Phi=\Phi_0/2$ the resistance $(R)$ exhibited a
maximum but the value of $R$ was a factor three less than the
normal state resistance $R_n$. If the cylinder has a nonuniform
thickness the ODV state can be realized as a stable state when
$\Phi=\Phi_0/2$ and it will lead to a resistive (but
superconducting) state even for very small currents and naturally
the resistance of such a state will be less than $R_n$ as was
found by Liu et al \cite{Liu}.

In conclusion, we studied the nonuniform state in a
superconducting ring in which the order parameter vanishes at one
point. It was shown that this state is characterized by a jump in
the phase of the order parameter by $\pi$ near the point $\psi=0$.
It allows, in correspondence with the ordinary two-dimensional
Abrikosov vortex, to call such a state a one-dimensional vortex.
This state is unstable in an uniform ring. In case of a nonuniform
ring (with variations of the geometrical ($d$, $w$) or physical
parameters ($\xi$, $\lambda$) along the ring) or for a ring with
an attached wire, this state may be realized in practise. For
rings with non-zero width the ODV state transforms into the usual
two-dimensional vortex. We also showed that this state is the
remnant of the saddle point state connecting two superconducting
states with different vorticity as found in a uniform ring. The
latter state becomes stable in a nonuniform ring.

The work was supported by the Flemish Science Foundation (FWO-Vl),
the "Onderzoeksraad van de Universiteit Antwerpen," the
"Interuniversity Poles of Attraction Program - Belgian State,
Prime Minister's Office - Federal Office for Scientific, Technical
and Cultural Affairs," and the European ESF-Vortex Matter. One of
us (D.Y.V.) is supported by a postdoctoral fellowship of FWO-Vl
and partially by a RFBR grant N01-02-16593. Discussions with Prof.
A. Geim and comments from Prof. J. Berger are gratefully
acknowledged.


\begin{references}
\bibitem{comm1}Numerical analysis shows that for large rings $L\gg \pi$ the solution
$t(s)=t_1-(t_1 -t_0)/{\rm cosh}^2(s\sqrt{(t_1-t_0)/2})$ found in
\cite{Langer} works very well for arbitrary currents $j<j_c$. The
reason is that for rings with $R \gg 1$ the "energy" $E$ is close
to $E_{max}$ and the roots $t_1$ and $t_2$ practically coincide.
As a result the situation is analogous to the one considered in
Ref. \cite{Langer}.
\bibitem{comm2} This result is a generalization of a result obtained
in Ref. \cite{Langer} for the saddle-point of Eq. (2) for the case
of an infinite long wire with vanishing transport current.
\bibitem{comm3} It is interesting to note that the results shown in
Fig. 2 correspond qualitatively to the time evolution of the order
parameter for an uniform ring in the case the vorticity increases
with one unit. The only difference is that for an uniform ring the
current density, generally speaking, is not equal to zero at the
point where the order parameter vanishes. The reason is that for
time-dependent processes the full current density is equal to the
sum of the superconducting $j_s$ and the normal $j_n$ parts. In
the point $f=0$, $j_s=0$ but $j_n\neq 0$.
\end{references}
\end{document}